\documentclass{JHEP3}
\usepackage{xspace}
\usepackage{amsmath}
\usepackage{epsfig}
\def\d{\textrm{d}}
\def\ca{\ensuremath{C_{\!\!A}}\xspace}
\def\as{\ensuremath{\alpha_s}\xspace}
\title{Energy Consumption and Jet Multiplicity from the Leading Log BFKL
  Evolution}
\author{Jeppe R.~Andersen\\DAMTP, Centre for Mathematical Sciences\\
  Wilberforce Road\\CB3 0WA\\Cambridge, UK\\and\\Cavendish Laboratory,
  University of Cambridge\\Madingley Road\\CB3 0HE\\Cambridge, UK}
\author{W.J.~Stirling\\IPPP, University of Durham\\South Road\\DH1
  3LE\\Durham, UK}
\abstract{We study the associated jet multiplicity arising from
  $t$--channel BFKL gluon evolution in forward dijet
  production at hadron colliders. Previous results have shown that the
  effect of conserving
overall energy and momentum is to introduce a pdf suppression that
completely compensates the predicted exponential BFKL rise with
rapidity difference between the leading dijets. However, we show
that there is still expected to be a significant amount of BFKL
radiation, especially in the central region, and we give
predictions for the multiplicity of the resulting mini--jets at
the LHC.}

\preprint{Cavendish
HEP--2002--21\\DAMTP--2002--154\\IPPP/02/80\\DCPT/02/160}
\begin{document}

\section{Introduction}
\label{sec:introduction}
The leading log BFKL
formalism~\cite{Kuraev:1976ge,Kuraev:1977fs,Balitsky:1978ic} resums large
logarithms in QCD processes with two large but different energy scales.  In
dijet production at hadron colliders the resummation is relevant in the
forward region~\cite{Mueller:1987ey}, when $\hat s\gg|\hat t|$ with $\hat s$
the parton centre of mass energy and $\hat t$ the square of the momentum
transfer.
This process has been promoted as the cleanest case at hadron colliders for
studying the BFKL evolution of the dominating $t$--channel gluon exchange,
since other processes where the BFKL formalism is applicable, for example the
DIS structure functions at low $x$, are plagued by an interplay between
perturbative and non--perturbative effects.

One of the most striking parton--level prediction of the leading
log BFKL formalism is an exponential rise in the dijet cross
section with increasing rapidity span of the dijets. However, as
has been pointed
out~\cite{Orr:1997im,DelDuca:1995ng,Andersen:2001kt}, this BFKL
enhancement may not be visible in  the hadronic cross section.
This is because the exponential rise of the cross section relies
on the emission of gluons from the BFKL evolution of the
$t$--channel gluon exchange, but the suppression from the parton
distribution functions (pdfs) for increasing partonic centre of
mass energy $\hat s$, at either the Tevatron or LHC, more than
compensates for the exponential rise. While this is certainly true
for dijet production, which is driven by the steeply falling (in
$x$) gluon pdf, recent results~\cite{Andersen:2001ja} suggest that
for processes depending instead on the much flatter (in the
relevant region) {\it valence quark} distributions, the BFKL
evolution might indeed lead to an increase in the hadron--level
cross section over the LO result.

Since the exponential rise in the cross section as a function of
the rapidity span of the dijets cannot therefore be considered a
precision BFKL prediction, other signatures of the BFKL evolution
have been proposed. In particular, the weakening of the angular
correlation of the leading dijets with increasing rapidity span,
as predicted by BFKL at the partonic level, is still present after
the convolution with the pdfs~\cite{Orr:1997im,DelDuca:1995fx}.
This angular decorrelation is a result of the multiple emission of
gluons, strong ordered in rapidity but unordered in transverse
momentum, resulting from the BFKL evolution. It is also
interesting to study the {\it jet multiplicity} resulting from the
BFKL evolution, since this could be another experimentally
verifiable signature of the theory. Such studies have been
undertaken
previously~\cite{DelDuca:1993zw,Forshaw:1998uq,Ewerz:1999tt,Ewerz:1999fn},
but these did not take into account the impact of the BFKL gluon
energies on the parton luminosities. In this paper we present a
study which does include this effect and thereby conserves overall
energy and momentum.

We will report on a study of the jet multiplicity from the BFKL
evolution of the $t$--channel gluon exchange in dijet production
at hadron colliders. We will also examine the energy of the
multi--jet events, in order to understand better why the
exponential rise of the partonic dijet cross section evidently
comes at too big a price in total energy to survive the
convolution with the pdfs. Both of these studies are performed
using the BFKL Monte Carlo approach of Ref.~\cite{Orr:1997im}. The
paper is organised as follows: we first briefly review BFKL
applied to dijet production and the technique of the BFKL MC in
Section~\ref{sec:bfkl-monte-carlo}, before reporting on the
results obtained for the total energy of the partonic BFKL events
(Section~\ref{sec:energy-cons-bfkl}). In
Section~\ref{sec:jet-mult-bfkl} we calculate the jet multiplicity
characteristic of BFKL evolution in dijet production, both in the
partonic and the hadronic case, and finally we present our
conclusions in Section~\ref{sec:conclusions}.

\section{Dijet Production at Hadron Colliders}
\label{sec:bfkl-monte-carlo}

In the high energy limit of $\hat s\gg|\hat t|\gg0$, the
gluon--gluon scattering cross section to leading order in $\ln\hat
s/|\hat t|$ but summed to all orders in $\alpha_s$ is given
by~\cite{Mueller:1987ey}
\begin{equation}
  \label{eq:dsigdptHE}
  \frac{\d \hat \sigma_{gg}(\Delta y)}{\d^2\mathbf{p}_{a\perp}\d^2
    \mathbf{p}_{b\perp}}=
  \left(\frac{\ca
      \as}{p_{a\perp}^2}\right)f
  (\mathbf{p}_{a\perp},-\mathbf{p}_{b\perp},\Delta y)
  \left(\frac{\ca \as}{p_{b\perp}^2}\right),
\end{equation}
where $\mathbf{p}_{a\perp}$ ($\mathbf{p}_{b\perp}$) is the
transverse momentum of the most forward (backward) jet, and
$\Delta y$ is the rapidity difference between them. The function
$f (\mathbf{p}_{a\perp},-\mathbf{p}_{b\perp},\Delta y)$ resums the
logarithms in $\hat s/|\hat t|$ arising from both virtual
corrections to, and rapidity ordered emission from, the
$t$--channel gluon exchange, and so solves the BFKL equation.
Although an analytic form for $f$ can be obtained, and the result
for the gluon--gluon scattering including the resummation of the
BFKL logarithms in Eq.~(\ref{eq:dsigdptHE}) thereby also solved
analytically, such an approach will potentially pose a problem
when it comes to calculating the hadronic cross section. This is
because in order to obtain an analytic solution, the gluons
emitted from the BFKL evolution cannot be counted in the
contribution to the parton momentum fractions, and therefore the
BFKL gluons that lead to the exponential rise in the partonic
cross section are emitted at no cost in energy. Although these
energy--conserving contributions to the centre of mass energy are
formally subleading compared to the contribution from the leading
dijets, they can have a huge impact on the parton distribution
functions and therefore on the normalisation of the hadronic cross
section\cite{Stirling:1994zs}, since the pdfs (and in particular
the gluon pdfs) are decreasing very rapidly in the relevant
region. The reformulation of the solution to the BFKL equation in
terms of an explicit sum and integration over the emitted gluons
and their rapidity ordered phase space was devised to solve this
problem\cite{Orr:1997im,Schmidt:1997fg}. In particular one finds
for the solution to the leading log BFKL equation\cite{Orr:1997im}
\begin{align}
f(\mathbf{q}_{a\perp},\mathbf{q}_{b\perp},\Delta y)\, =  \,
  \sum_{n=0}^{\infty} f^{(n)}(\mathbf{q}_{a\perp},\mathbf{q}_{b\perp},\Delta y) \; .
\label{eq:b7}
\end{align}
where we have set $\mathbf{q}_{a\perp}=\mathbf{p}_{a\perp},
\mathbf{q}_{b\perp}=-\mathbf{p}_{b\perp}$ and
\begin{align}
  \begin{split}
    f^{(0)}(\mathbf{q}_{a\perp},\mathbf{q}_{b\perp},\Delta y) &= \left[
      \frac{\mu^2}{q_{a\perp}^2} \right]^{\bar\alpha_s\Delta y}\,
    \,\frac{1}{2}\, \delta^{(2)} (\mathbf{q}_{a\perp}-\mathbf{q}_{b\perp} )\,,
    \\
    f^{(n\geq 1)}(\mathbf{q}_{a\perp},\mathbf{q}_{b\perp},\Delta y) &=
    \left[ \frac{\mu^2}{q_{a\perp}^2} \right]^{\bar\alpha_s\Delta y}\,
    \left\{ \prod_{i=1}^{n} \int \d^2 \mathbf{k}_{i\perp}\, \d y_i \, {\cal F}_i
    \right\} \,\frac{1}{2}\, \delta^{(2)} (\mathbf{q}_{a\perp}-\mathbf{q}_{b\perp}
    - \sum_{i=1}^n \mathbf{k}_{i\perp})\,,
    \\
    {\cal F}_i &= \frac{\bar\alpha_s}{\pi k_{i\perp}^2}\,
    \theta(k_{i\perp}^2 -\mu^2)\, \theta(y_{i-1}-y_i)\, \left[ {
        \frac{(\mathbf{q}_{a\perp} +\sum_{j=1}^{i-1}\mathbf{k}_{j\perp} )^2}
        {(\mathbf{q}_{a\perp} +\sum_{j=1}^{i}\mathbf{k}_{j\perp} )^2 }}
    \right]^{\bar\alpha_s y_i}\,,
\label{eq:b8}
\end{split}
\end{align}
with $\bar\alpha_s=\ca\alpha_s/\pi$\footnote{This is the fixed $\alpha_s$
  result. The solution for running $\alpha_s$ is only slightly more
  complicated and is given for example in Ref.~\cite{Orr:1997im}. In the
  numerical results that follow we use a fixed coupling with a value of
  $\alpha_s(20\mathrm{GeV})=0.1635$.} and $\mu$ the resolution scale of the
Monte Carlo. For small $\mu$ the sum in Eq.~(\ref{eq:b7}) is only weakly
dependent on $\mu$. This Monte Carlo formulation has been applied to studies
of the dijet production rate and angular decorrelation at large rapidity
separation at hadron colliders.  One finds that the decrease in parton flux
as a result of the increased centre of mass energy when the BFKL radiation is
taken into account more than compensates for the BFKL exponential rise in the
partonic cross section. The details of this effect obviously depends on the
specific shape of the pdfs.  In this paper we will therefore first study
directly the impact of the BFKL radiation on the centre of mass energy for
gluon--gluon scattering.

\section{Energy Consumption of the BFKL evolution}
\label{sec:energy-cons-bfkl} Using the solution of the BFKL
equation in the form of Eqs.~(\ref{eq:b7},\ref{eq:b8}) together
with the Monte Carlo implementation of the integrations, we can
answer the question of how much energy goes into creating the LL
BFKL radiation. When energy and momentum conservation is applied,
the parton momentum fractions are given by
\begin{align}
  \begin{split}
    \label{eq:fullbjorkenx}
  x_a=&\frac{p_{a\perp}}{\sqrt s}e^{y_a} +
  \sum_{i=1}^{n}\frac{k_{i\perp}}{\sqrt s}e^{y_i}+\frac{p_{b\perp}}{\sqrt s}e^{y_b}\\
  x_b=&\frac{p_{a\perp}}{\sqrt s}e^{-y_a} +
  \sum_{i=1}^{n}\frac{k_{i\perp}}{\sqrt s}e^{-y_i}+\frac{p_{b\perp}}{\sqrt s}e^{-y_b},
\end{split}
\end{align}
with the overall centre of mass energy squared given by $\hat
s=x_ax_bs$ where $\sqrt{s}$ is the energy of the hadron collider.

\begin{figure}[htbp]
  \centering
  \epsfig{width=10cm,file=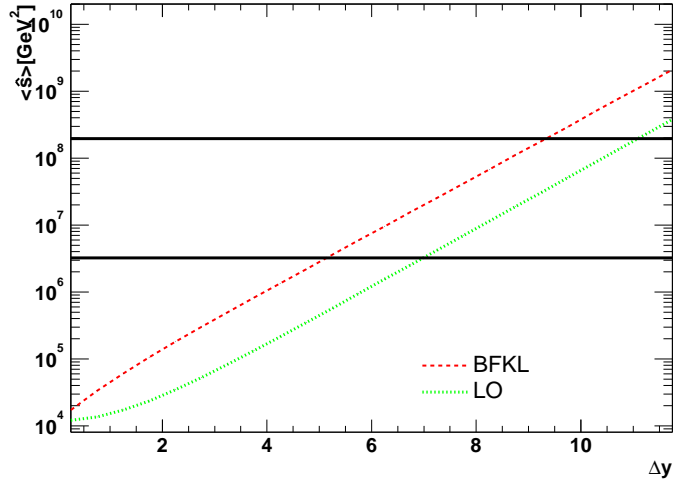}
  \caption{The average centre
    of mass energy in $gg\to gg$ scattering with (red/dashed) and without
    (green/dotted) 
    BFKL evolution of the $t$ channel gluon, with
    $p_{\perp\mathrm{min}}=20$~GeV for the dijets and $\alpha_s=0.1635$. Also
    plotted is the hadronic centre of mass energy squared for the Tevatron
    ($(1.8\mathrm{TeV})^2$) and the LHC($(14\mathrm{TeV})^2$).}
  \label{fig:avghats}
\end{figure}
In Fig.~\ref{fig:avghats} we show the average centre of mass
energy squared for dijet production in fixed leading order QCD
(green/dotted) and for dijet production in the high--energy limit
with BFKL evolution of the $t$--channel gluon exchange
(red/dashed). The prediction for pure dijet production is
indistinguishable in this plot from the `standard' prediction for
BFKL evolution, when the BFKL equation is solved analytically and
the contribution from the BFKL gluons to the centre of mass energy
is neglected.

The contribution from the BFKL radiation to the centre of mass
energy is formally subleading compared to the contribution from
the leading dijets. Indeed we see in Fig.~\ref{fig:avghats} that
asymptotically (which evidently is reached quickly), the two
curves have the same slope, even though they are offset by about
1.5 units of rapidity. This means that the kinematic limit of
dijet production with BFKL evolution at hadron colliders is
reached about 1.5 units of rapidity before an estimate based on
the energy of the leading dijets only. We have also indicated the
hadronic centre of mass energy squared for the Tevatron
((1.8TeV)$^2$) and the LHC ((14TeV)$^2$). The rapid decrease of
the pdfs as the kinematic limit is approached means that the
horizontal lines of interest for a given hadron collider lie
considerably lower than the lines indicated on the figure.
Therefore the effect of including the contribution from the BFKL
radiation in the overall energy and momentum conservation is
larger than may appear at first glance. When considering the
implications on the cross section, it must also be remembered that
the BFKL evolution predicts an exponential rise with the rapidity
span $\Delta y$, i.e. $\hat{\sigma} \sim \exp(\lambda\Delta y)$.
Therefore effectively reducing the available rapidity span has a
sizeable impact on the prediction for the cross section.

We can see explicitly why the two curves in Fig.~\ref{fig:avghats}
have the same asymptotic slope. As we will see later, the
radiation from the BFKL chain is distributed evenly in rapidity
along the chain. We can approximate the Bjorken $x$'s given by
Eq.~(\ref{eq:fullbjorkenx}) by assuming that all the $k_{\perp i}$
are equal to $k$ and that the $n$ BFKL gluons are distributed
evenly over the rapidity span $\Delta y$, separated by $\delta y$
such that $\Delta y = (n+1) \delta y$. The Bjorken $x$'s for the
$2\to 2 + n$ scattering then become (with $z=e^{-\delta y}$ and
assuming that $y_0 = -\Delta y/2,\; y_{n+1} = + \Delta y/2$ ---
the centre of mass energy is independent of this assumption)
\begin{align}
  \label{eq:Bjorkenxeven}
  x_a=x_b=\frac k {\sqrt s} e^{\Delta y/2}(1+z+z^2+\cdots+z^n)=\frac k {\sqrt s}
  e^{\Delta y/2} \frac{1-z^{n+1}}{1-z}.
\end{align}
In the large $\Delta y$ limit, with $n\to\infty$ for evenly spaced
radiation, we find
\begin{align}
  \label{eq:naiveresumenergy}
  \hat s\propto k^2e^{\Delta y}\frac 1 {(1-e^{-\delta y})^2},
\end{align}
to be compared with the pure dijet prediction $\hat s \propto
k^2e^{\Delta y}$, which has the same dependence on $\Delta y$. It
is radiation from the region of the chain close to the endpoints
that contributes  most to $\hat s$, since the middle part of the
chain will give exponentially suppressed contributions to the
energy (this is just a refinement of the asymptotic argument for
dropping the contribution from the chain altogether). This
explains why asymptotically there is only a difference in the
normalisation and not the shape of the two curves in
Fig.~\ref{fig:avghats}. From Eq.~(\ref{eq:naiveresumenergy}) we
see that the smaller the $\delta y$, the bigger the difference in
normalisation. Note that a smaller $\delta y$ can be achieved by
increasing \as, thereby increasing the amount of BFKL radiation
everywhere, and specifically also in the region close to the
endpoints of the BFKL chain.

The observation that there is insufficient energy available at
present--day colliders for all the BFKL radiation resummed in the
analytic approach to be emitted without penalty motivated the
introduction in Ref.~\cite{DelDuca:1995ng} of a `reduced effective
rapidity separation', to be used when making phenomenological
predictions of BFKL signatures for comparison with data. The idea
behind this is to emulate the reduction of phase space for BFKL
radiation, dictated by energy and momentum conservation, by reducing
the rapidity span $\Delta y$ that is used in the solution to the
BFKL equation.


\section{Jet Multiplicity of the BFKL Evolution}
\label{sec:jet-mult-bfkl} In this section we will study the jet
multiplicity from the leading log BFKL evolution, first for
partonic gluon--gluon scattering and then for the full hadronic
cross section when proper account is taken of the energy and
momentum carried by the BFKL gluons.  The parton multiplicity, as
predicted from leading log BFKL evolution, is equivalent (up to
sub--leading terms) to the result obtained in the CCFM approach,
in which colour coherence effects are taken into
account~\cite{Forshaw:1998uq,Salam:1999ft}. This supports the
correspondence between gluons emitted from the BFKL chain and jets
appearing in the detector.

\subsection{Mini--jet Parton Multiplicities}
\label{sec:mini-jet-mult}

In this section we study the mini--jet multiplicity from the BFKL
chain. A mini--jet is here defined as a gluon with transverse
momentum $q_{i\perp}<\mu_m$, where $\mu_m$ is the maximum allowed
transverse momentum of a mini--jet, and the scales of the problem
are ordered according to
\begin{align}
  \label{eq:momentumordering}
\mu<q_{i\perp}<\mu_m\ll q_{a,b\perp} .
\end{align}
Recall that $\mu$ is the resolution scale of the BFKL MC solution
in Eq.~(\ref{eq:b8}). Defined in this way, the sum over
$n$--mini--jet rates with a resolution scale $\mu$ and upper scale
$\mu_m$ should equal the 0--jet rate with a resolution scale of
$\mu_m$ (up to terms of order $\mu^2/\mu_m^2$). Introducing a
cut--off in the transverse momentum integral in the definition of
$f^{(1)}(\mathbf{q}_{a\perp},\mathbf{q}_{b\perp},\Delta y)$ in
Eq.~(\ref{eq:b8}) and denoting the function with constrained
integration
$f^{(1)}_c(\mathbf{q}_{a\perp},\mathbf{q}_{b\perp},\Delta y)$, we
find
\begin{align}
  f^{(1)}_c(\mathbf{q}_{a\perp},\mathbf{q}_{b\perp},\Delta y)=&
\left[ \frac{\mu^2}{q_{a\perp}^2} \right]^{\bar\alpha_s\Delta y} \int \d^2
\mathbf{k}_{1\perp}\,\int_0^{\Delta y} \d y_1 \frac 1 2 \delta^{(2)} (\mathbf{q}_{a\perp}-\mathbf{q}_{b\perp}
    - \mathbf{k}_{1\perp}) \\
&\theta(k_{1\perp}^2 -\mu^2)\,\theta(\mu^2_m-k_{1\perp}^2)\frac{\bar\alpha_s}{\pi k_{1\perp}^2}
\, \left[ {
        \frac{(\mathbf{q}_{a\perp})^2}
        {(\mathbf{q}_{a\perp} +\mathbf{k}_{1\perp} )^2 }}
    \right]^{\bar\alpha_s y_1}.
\end{align}
If we approximate the term in the square brackets by unity (by
virtue of the applied ordering of momenta in
(\ref{eq:momentumordering})) we find that
$f^{(1)}_c(\mathbf{q}_{a\perp},\mathbf{q}_{b\perp},\Delta y)$ can
be approximated by
\begin{align}
  f^{(1)}_c(\mathbf{q}_{a\perp},\mathbf{q}_{b\perp},\Delta y)\approx
\left[ \frac{\mu^2}{q_{a\perp}^2} \right]^{\bar\alpha_s\Delta y} \frac 1 2
\delta^{(2)} (\mathbf{q}_{a\perp}-\mathbf{q}_{b\perp})\, \Delta y\,
\bar\alpha_s\ln\frac{\mu_m^2}{\mu^2},
\end{align}
where we have set $\delta^{(2)} (\mathbf{q}_{a\perp}-\mathbf{q}_{b\perp} -
\mathbf{k}_{1\perp})\approx\delta^{(2)}
(\mathbf{q}_{a\perp}-\mathbf{q}_{b\perp})$. Similarly, one finds
\begin{align}
  f^{(n)}_c(\mathbf{q}_{a\perp},\mathbf{q}_{b\perp},\Delta y)\approx
\left[ \frac{\mu^2}{q_{a\perp}^2} \right]^{\bar\alpha_s\Delta y} \frac 1 2
\delta^{(2)} (\mathbf{q}_{a\perp}-\mathbf{q}_{b\perp})\, \frac{\Delta y^n}{n!}\,
\left(\bar\alpha_s\ln\frac{\mu_m^2}{\mu^2}\right)^n.
\end{align}
Therefore the contribution to the sum
$f(\mathbf{q}_{a\perp},\mathbf{q}_{b\perp},\Delta y)$ in
Eq.~(\ref{eq:b7}) with no gluon emission with transverse momentum
squared greater than $\mu_m^2$ calculated this way is
\begin{align}
  \sum_nf^{(n)}_c(\mathbf{q}_{a\perp},\mathbf{q}_{b\perp},\Delta y)\approx&
  \left[ \frac{\mu^2}{q_{a\perp}^2} \right]^{\bar\alpha_s\Delta y} \frac 1 2
\delta^{(2)} (\mathbf{q}_{a\perp}-\mathbf{q}_{b\perp})\sum_{n=0}^\infty\frac
  1 {n!}\left(\Delta
  y\,\bar\alpha_s\,\ln\frac{\mu_m^2}{\mu^2}\right)^n\nonumber\\
=&\left[ \frac{\mu^2}{q_{a\perp}^2} \right]^{\bar\alpha_s\Delta y} \frac 1 2
\delta^{(2)}
  (\mathbf{q}_{a\perp}-\mathbf{q}_{b\perp})\exp\left(\Delta
  y\,\bar\alpha_s\,\ln\frac{\mu_m^2}{\mu^2} \right)\\\nonumber
=&\left[
      \frac{\mu_m^2}{q_{a\perp}^2} \right]^{\bar\alpha_s\Delta y}\,
    \,\frac{1}{2}\, \delta^{(2)} (\mathbf{q}_{a\perp}-\mathbf{q}_{b\perp} ),
  \label{eq:nojetrate}
  \end{align}
which we recognise as
$f^{(0)}(\mathbf{q}_{a\perp},\mathbf{q}_{b\perp},\Delta y)$
evaluated with a resolution scale $\mu_m$. This serves as a check
of the consistency of the physical picture emerging from the Monte
Carlo solution to the BFKL equation.

\subsection{Partonic Jet Multiplicities}
\label{sec:part-hadr-jet} It proves much harder to obtain analytic
predictions for multi--jet multiplicities or predictions for the
rates of harder mini--jets from the BFKL chain, i.e. with the
strong ordering $\mu_m\ll q_{a,b\perp}$ constraint relaxed. The
results which follow are therefore obtained using the Monte Carlo
approach. In all of these calculations we have chosen a resolution
scale for the Monte Carlo implementation of
Eqs.~(\ref{eq:b7},\ref{eq:b8}) of $\mu=1$~GeV and a fixed value
for the coupling $\alpha_s=0.1635$.
\begin{figure}[tb]
  \centering
  \epsfig{width=10cm,file=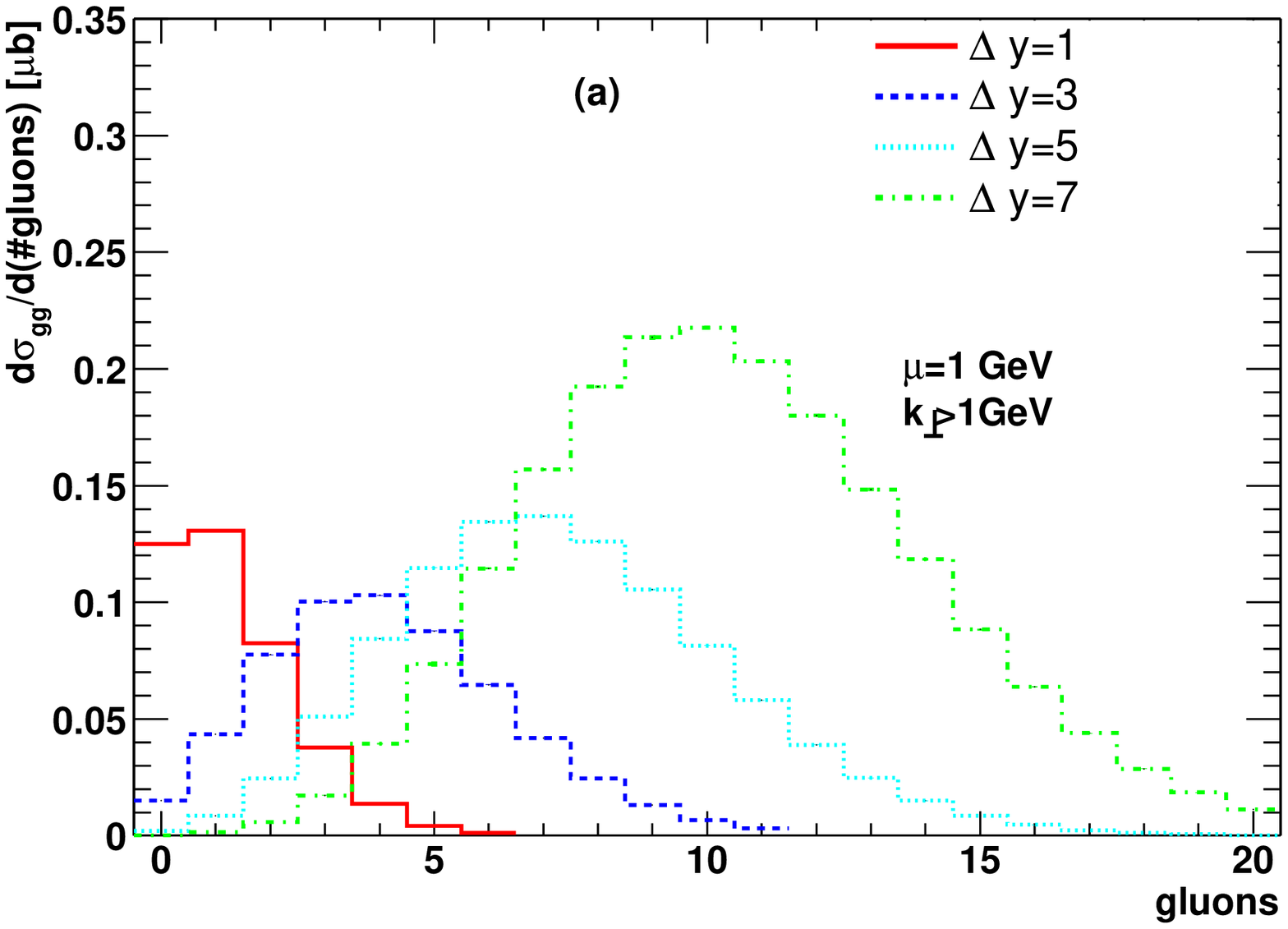}
  \epsfig{width=10cm,file=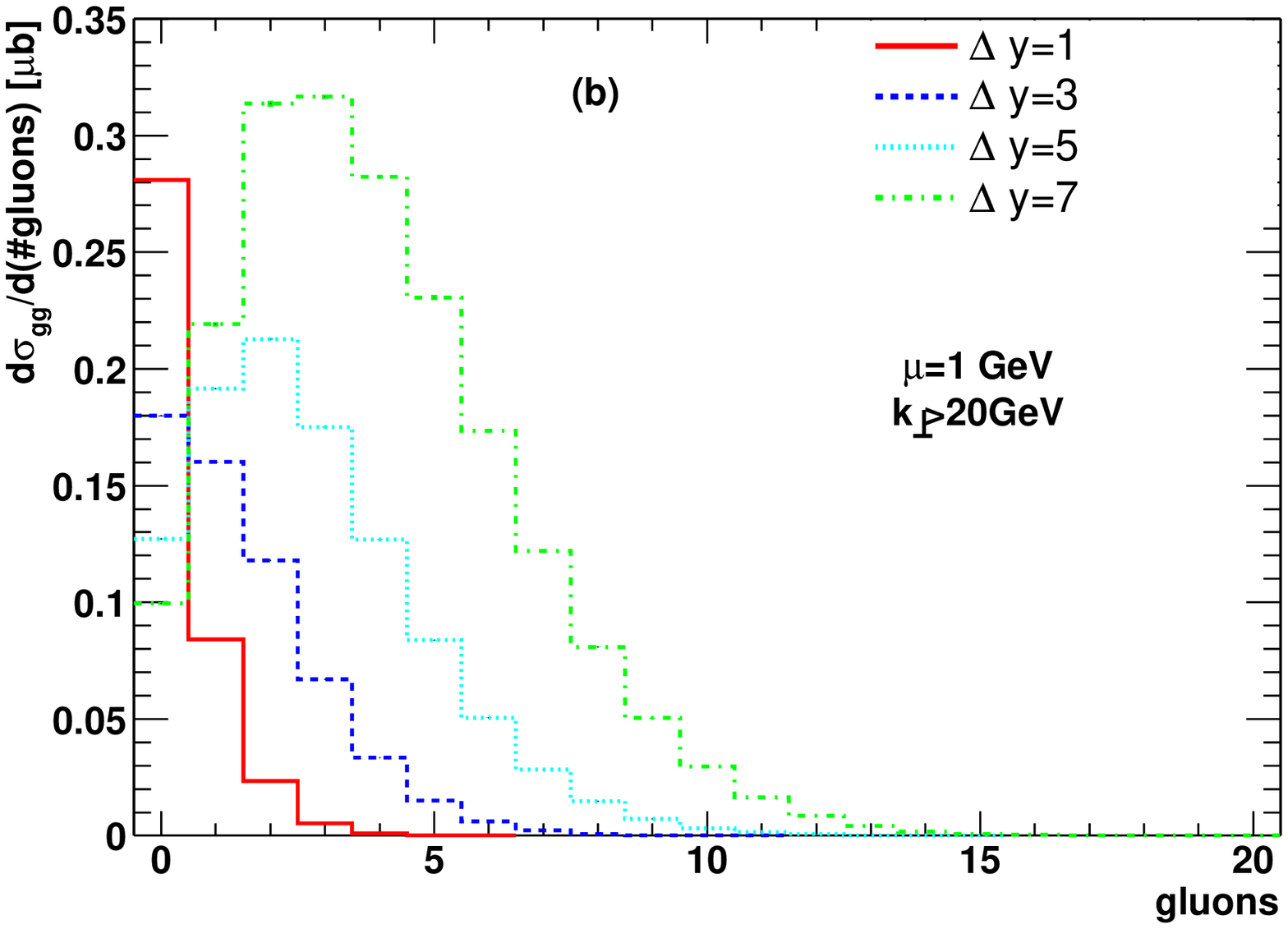}
  \caption{The contribution to the partonic cross section for choices of the
  rapidity separation from different numbers of resolved
  gluons with (a) $k_{i\perp}>1$~GeV and (b) $k_{i\perp}>20$~GeV. The leading dijets
  have $p_{\perp\mathrm{min}}=20$~GeV in both cases.}
  \label{fig:resolvedgluons}
\end{figure}
In Fig.~\ref{fig:resolvedgluons}(a) we plot the contribution to
the partonic cross section from  different numbers of resolved
(i.e. $k_{i\perp}>\mu=1$~GeV) gluons from the BFKL chain for a
selection of rapidity spans of the chain. In calculating the
partonic cross section (Eq.~(\ref{eq:dsigdptHE})) we have chosen a
cut--off on the minimum transverse momentum of the jets $a,b$ of
20~GeV. Such a simple cut--off, in which the threshold is the same
for jet $a$ and jet $b$, is known to cause incomplete
cancellations of virtual and real IR divergences in a full NLO QCD
calculation~\cite{Frixione:1997ks}, but as was shown in
Ref.~\cite{Andersen:2001kt} this effect is not present in the case
of BFKL evolution. This is essentially because that at NLO the
transverse momentum of the third gluon is by momentum conservation
determined by the two others, and so the collinear (and soft)
phase space of the third gluon can be restricted by cuts in the
allowed phase space for the two harder jets. However, when more
gluons are emitted (e.g.~by BFKL evolution) this is no longer
true, and the extra gluons can populate the IR regions
irrespective of the cuts on the harder jets. We can therefore
safely choose just a simple cut--off for our calculation.

The cross section for a given value of the rapidity difference
between the leading dijets is given by the integral of the curves
in Fig.~\ref{fig:resolvedgluons}(a). If one were to plot the
average number $\langle n \rangle$ of emitted BFKL gluons as a
function of the rapidity span of the BFKL chain, one would find
that  $\langle n \rangle$  increases linearly with the rapidity
span. This feature will be evident in later plots.

Fig.~\ref{fig:resolvedgluons}(b) contains the same curves as in
(a), but now for harder BFKL gluons with $k_{i\perp}>20$~GeV.
These latter curves are obtained by running the Monte Carlo with a
resolution scale $\mu_r=1$~GeV and then counting how many gluons
with $k_{i\perp}>20$~GeV the event contains. These plots are for
the partonic cross section, and so these harder gluons from the
BFKL chain contribute significantly to the increase in the average
centre of mass energy seen in Fig.~\ref{fig:avghats} for BFKL in
comparison with the LO prediction.

\begin{figure}[tbh]
  \centering
  \epsfig{width=10cm,file=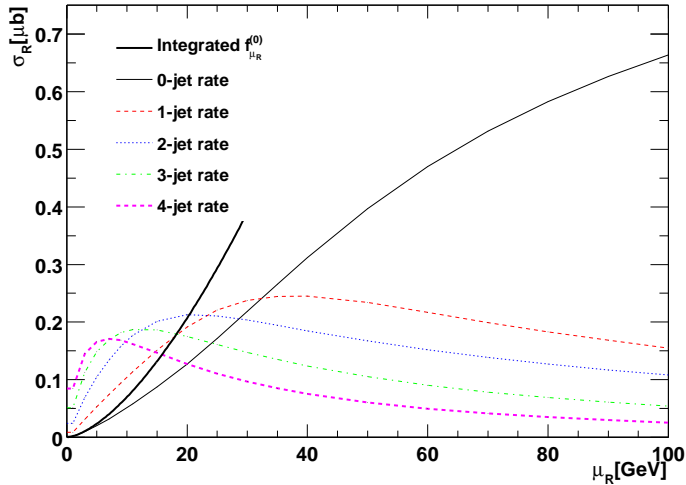}
  \caption{The $0$--, $1$, $2$--, $3$--
    and $4$--jet parton--level cross sections as a function of $\mu_R$, for a
    rapidity span of $\Delta y = 5$ and $p_{\perp\mathrm{min}}=20$~GeV for
    the leading dijets.  Also shown is the analytic $0$--jet prediction valid
    for small $\mu_R$. }
  \label{fig:zeroonejet}
\end{figure}
Fig.~\ref{fig:resolvedgluons} also shows that the jet multiplicity varies
with the minimum $k_{i\perp}$ in the expected way. This
is also seen in Fig.~\ref{fig:zeroonejet}, where the $0$--, $1$--, $2$--,
$3$-- and $4$--jet cross sections are shown as a function of $\mu_R$ for a
rapidity span of $\Delta y = 5$ and $p_{\perp\mathrm{min}}=20$~GeV for the
leading dijets. Note that for $\mu_R\approx p_{\perp\mathrm{min}}$ the
multijet cross sections are all of similar magnitude (for the relatively
small number of jets considered here). Also shown in
Fig.~\ref{fig:zeroonejet} is the analytic prediction for the zero jet rate,
valid for $ \mu_R \ll p_{\perp\mathrm{min}}$, obtained by integrating
Eq.~(\ref{eq:nojetrate}) over $q_{a,b\perp}$:
\begin{align}
\sigma_R=\frac{\alpha_s^2 C_A^2\pi}{2 p_{\perp\mathrm{min}}^2}
\frac{1}{1+\bar\alpha_s\Delta y} \left(
\frac{\mu_R^2}{p_{\perp\mathrm{min}}^2}
\right)^{\bar\alpha_s\Delta y}.
\end{align}

\begin{figure}[htbp]
  \centering
  \epsfig{width=10cm,file=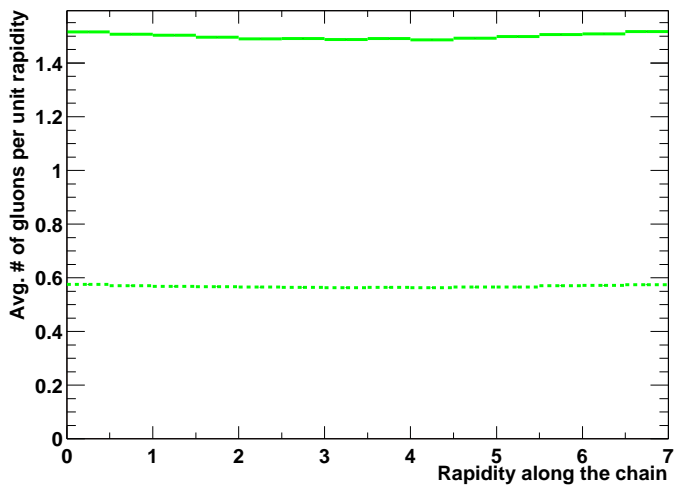}
  \caption{The average density of emitted gluons along the BFKL chain. Please
  see text for further details.}
  \label{fig:avgdensgluons}
\end{figure}
In Fig.~\ref{fig:avgdensgluons} we show the average density in
rapidity of the gluons emitted from the BFKL chain (not counting
the two leading jets) in the case of gluon--gluon scattering for a
gluon chain spanning 7 units of rapidity. The two lines on the
plot correspond to the average density of resolved BFKL gluons
($k_{i\perp}>1$~GeV) and harder gluons ($k_{i\perp}>20$~GeV). The
density of gluon emission along the chain is the observable best
suited for illustrating the effects of taking into account the
energy of this radiation when calculating the parton momentum
fractions for the hadronic cross section. As expected from the
analytic approach, the density in rapidity of emitted gluons is
(relatively) constant along the chain.

\subsection{Hadronic Jet Multiplicities}
\label{sec:hadr-jet-mult} If the contributions from the BFKL gluon
radiation to the parton momentum fractions in
Eq.~(\ref{eq:fullbjorkenx}) are neglected, the parton--level
result of an exponential growth (over the LO result) in cross
section when BFKL evolution is taken into account obviously
carries through to the hadronic cross section. As has been
discussed before (see for example
~Refs.~\cite{Orr:1997im,DelDuca:1995ng}), this will no longer be
the case when energy and momentum conservation is imposed by
taking into account the BFKL gluon radiation in evaluating the
parton momentum fractions. This is because the dijet cross section
is driven by the gluon pdf, which falls off very sharply at medium
and high $x$. So despite the fact that the contribution of the
BFKL gluons to the parton momentum fractions is formally
subleading, the numerical impact for dijet production is large and
gets magnified by the sharply decreasing gluon pdf (see for
example ~Ref.~\cite{Stirling:1994zs} for arguments on the error in
the normalisation of the result when neglecting the contribution
of the BFKL gluons to the parton momentum fraction). The impact on
the pdfs counteracts the expected exponential rise in cross
section with growing rapidity difference between the leading
dijets~\cite{Orr:1997im}, and results in an almost no--change
situation for the cross section. The fine details obviously depend
on the exact shape of the pdfs, but the conclusions do not. In the
following we will use the fits to the pdfs of
Ref.\cite{Martin:1999ww}.

The observation that the shape of the differential cross section
as a function of rapidity is not expected to change dramatically
when including BFKL evolution of the $t$--channel gluon has led to
the study of other observables such as the azimuthal correlation
of the leading dijets\cite{Stirling:1994zs}, where the dependence
on the pdfs is expected to be less pronounced.

\begin{figure}[tbh]
  \centering
  \epsfig{width=10cm,file=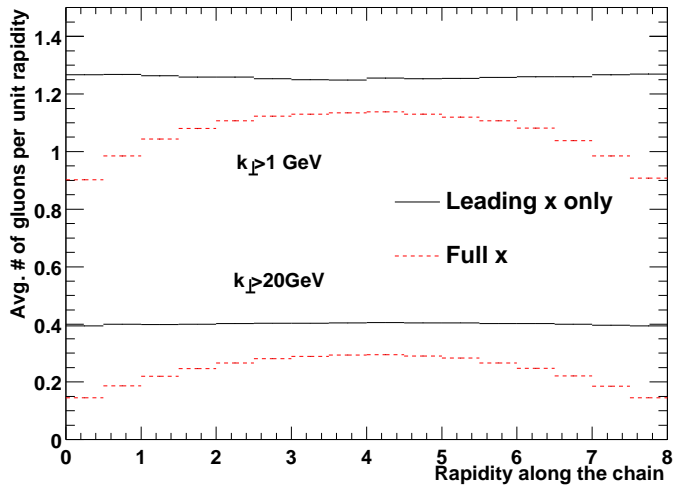}
  \caption{The
    average density of emitted gluons along the BFKL chain for the constant
    coupling formalism in the case hadronic dijet production. Please see text
    for further details.}
  \label{fig:avgjetspdf}
\end{figure}
Here we will study how energy and momentum conservation influences
the associated gluon (i.e.~jet) multiplicity for hadronic dijet
production at the LHC. If the effect of the energy and momentum
conserving constraint was to suppress the emissions of BFKL gluons
completely, then BFKL would be irrelevant at such collider
energies and fixed--order calculations would be adequate: there
simply would not be enough phase space available for the gluon
emission resummed through the BFKL equation. In
Fig.~\ref{fig:avgjetspdf} we plot the average density (in
rapidity) of gluons emitted from a BFKL chain with ends fixed at
rapidities $-3.5$ and $3.5$ in the case of the hadronic cross
section, i.e. with pdfs included. Just as for
Fig.~\ref{fig:avgdensgluons}, we plot the density of both resolved
($k_{i\perp}>1$~GeV) and harder ($k_{i\perp}>20$~GeV) gluons, and
we have chosen to plot the results obtained by two choices for
evaluating the parton momentum fractions, namely the full version
of Eq.~(\ref{eq:fullbjorkenx}) and the version where the BFKL
gluon contribution is ignored.

Let us start by discussing the two upper curves of Fig.~\ref{fig:avgjetspdf},
corresponding to the density of resolved gluons. The upper line is obtained
by ignoring the BFKL gluon contribution to the parton momentum fractions, and
so corresponds to the case where one obtains an exponential increase over the
LO cross section. The result for the gluon density is to be compared with the
upper line in Fig.~\ref{fig:avgdensgluons}.  We see that, just as expected,
the density of gluons is not modified significantly in this case\footnote{The
  small difference compared to Fig.~\ref{fig:avgdensgluons} is caused by the
  fact that the transverse momentum spectrum of the leading dijets is
  softened after convolution with the pdfs. The softening of the leading
  dijets results also in a softening of the BFKL radiation.}. This is needed
in order to maintain the exponential rise in cross section over the LO
result. However, when we include the contribution of the BFKL gluons to the
parton momentum fraction, the result is changed to the dashed (red) curve,
and we see that the density of radiated gluons is reduced, especially at the
ends of the chain where the effect on the energy consumption is exponentially
enhanced compared to the effect of radiation in the middle of the chain (in
this symmetric configuration).

The corresponding curves for harder gluons with
$k_{i\perp}>20$~GeV are also plotted in Fig.~\ref{fig:avgjetspdf},
and we see that in this case the relative effect of conserving
energy and momentum is larger, especially at the end points. This
is of course because the effect of energy and momentum
conservation is most severe for hard radiation near the end points
of the chain.

But the most important conclusion to be drawn from
Fig.~\ref{fig:avgjetspdf} is that the suppression of the BFKL
evolution is not as dramatic as suggested by the cancellation of
the BFKL rise in the dijet cross section by the pdf suppression.
Conservation of energy and momentum still leaves sufficient phase
space (at the LHC) to allow for a significant number of BFKL
gluons to be emitted, and therefore for the BFKL evolution to be
relevant. Therefore one cannot conclude from the predicted lack of
rise in the hadronic dijet cross section that BFKL evolution is
irrelevant. It is the abundance of BFKL radiation even when
applying the full parton momentum fractions that results in
sizeable angular decorrelations predicted between dijets at the
LHC\cite{Orr:1997im}.

Fig.~\ref{fig:avgjetspdf} also supports the idea of introducing an
effective reduced rapidity span of the BFKL chain in analytic
calculations. The BFKL evolution is obviously most important in a
reduced rapidity span\cite{DelDuca:1995ng}, where the BFKL
emission (serving as a measure of the importance of BFKL evolution
in the presence of energy and momentum conservation) is only
slightly reduced compared to the partonic prediction.

It is not clear {\it a priori} in which region of rapidities an
analytic prediction (ignoring energy and momentum conservation)
will be valid, since first ``asymptotic'' values of the rapidity
span have to be reached for the formalism to be valid, but these
``asymptotic values'' cannot be too large in order for the total
energy available at the collider not to restrict too much the
phase space available for BFKL gluon emission.  In essence,
examining this problem is what the construction of the BFKL Monte
Carlo approach is all about.

\section{Conclusions}
\label{sec:conclusions} We have examined the jet multiplicity
predicted from BFKL evolution of a $t$--channel gluon exchange in
dijet production, at both the partonic and hadronic levels, at the
LHC. Previous results have shown that the effect of conserving
overall energy and momentum at present and future hadron colliders
is to introduce a pdf suppression that completely compensates the
predicted exponential BFKL rise with rapidity difference between
the leading dijets. However, in the present analysis we have shown
that there is still predicted to be a significant amount of BFKL
radiation, and therefore that BFKL evolution {\it will} be
relevant for QCD observables less dependent on the parton
distribution functions. These include the angular decorrelation of
the leading dijets. Furthermore, this multi--jet emission will of
course provide a background to multi--jet observables in processes
beyond the Standard Model at hadron colliders.

\acknowledgments 
We are grateful to Bryan Webber and Gavin Salam for useful discussions.

\bibliographystyle{JHEP}
\bibliography{database}

\end{document}